\documentclass{epl}
\usepackage{graphicx}

\newcommand \be  {\begin{equation}}
\newcommand \bea {\begin{eqnarray} \nonumber}
\newcommand \ee  {\end{equation}}
\newcommand \eea {\end{eqnarray}}

\begin{document}

\title{Non-compact local excitations in spin glasses}

\shortauthor{J. Lamarcq {\it et al.} }
\author{J. Lamarcq$^1$, J.-P. Bouchaud$^1$, O. C. Martin$^2$ and M. M\'ezard$^2$}

\institute{
$^1$~Service de Physique de l'Etat Condens\'e,\\
Orme des Merisiers --- CEA Saclay,\\
91191 Gif sur Yvette Cedex FRANCE \\
$^2$~Laboratoire de Physique Th\'eorique et Mod\`eles Statistiques,
\\ B\^atiment 100 --- Universit\'e Paris Sud,\\
91405 Orsay FRANCE
}

\date{\today}
\pacs{02.60.Pn}{Numerical optimization}
\pacs{75.10.Nr}{Spin glass and other random models}

\maketitle

\begin{abstract}
We study numerically the local low-energy excitations in the $3d$
Edwards-Anderson model for spin glasses.  Given the ground state, we
determine the lowest-lying connected cluster of flipped spins with a
fixed volume containing one given spin. These excitations are not
compact, having a fractal dimension close to two, suggesting an
analogy with lattice animals. Also, their energy does {\it not} grow
with their size; the associated exponent is slightly negative whereas
the one for {\it compact} clusters is positive. These findings call
for a modification of the basic hypotheses underlying the droplet
model.
\end{abstract}

\section{Introduction}
In spite of over twenty years of attention, the nature of the spin
glass phase is still an open question~\cite{MarinariParisi99b}.  If
one is guided by the mean field theory of the infinite range spin
glass model \cite{MezardParisiVirasoro}, one is lead to a picture
where three dimensional ($3d$) spin glasses have many unrelated
valleys whose free-energies differ by $O(1)$.  Extending this to zero
temperature, one should be able to find large excitations above the
ground state which cost only $O(1)$ in energy, regardless of their
volume.  An important problem is to characterize these excitations
geometrically.  In the droplet picture~\cite{FisherHuse86}, the
lowest-lying excitations are postulated to be {\it compact}, {\it
i.e.}, their volume grows as the cube of their characteristic linear
size $\ell$.  In this picture, a spin glass is like a disguised
ferromagnet with just two (spin reversed) pure states. Furthermore,
the lowest-lying excitations have {\it typical} energies that grow as
a power of their size $\ell$, {\it i.e.}, as $\Upsilon \ell^\theta$
with a positive exponent $\theta$.  
Only in {\it rare} cases (the probability of which
decreases as $\ell^{-\theta}$) will a droplet energy be $O(1)$,
whereas the mean field picture suggests that this probability remains
of order one even for large $\ell$.  The exponent $\theta$ has been
estimated numerically to be about $0.2$ in $3d$ by measuring the
energy difference between ground states when applying periodic and
antiperiodic boundary conditions~\cite{BrayMoore84,McMillan84}.  With
this procedure, a ``domain wall'' of linear size $L$ is forced through
the sample and cuts it into two compact pieces.  Such an approach
implicitly assumes that a domain wall and the surface of a droplet are
topologically similar, in which case one may hope to identify the above
exponent $\theta$ describing {\it local} ({\it i.e.}, $\ell \ll L$)
excitations with an a priori different exponent $\theta_{dw}$ 
describing domain wall excitations (whose
characteristic size is that of the whole system).  Both in
two~\cite{Kawashima00,Ritort01} and in
three~\cite{KrzakalaMartin00a,PalassiniYoung00a,HedHartmann00a,MarPar01}
dimensions, there are now indications that this assumption is
incorrect.

This work considers the {\it local} excitations of the $3d$
Edwards-Anderson model~\cite{EdwardsAnderson75}.  Our goal is to test
the key ingredients of the droplet model: droplets are compact and
their energies grow with their characteristic size.  For our numerical
investigation, we construct the connected clusters of spins of lowest
energy that contain a specified number of spins and a given site,
hereafter called ``minimum energy clusters'' (\textsc{mec}).
We focus on the statistical properties of the energy and
geometry of a \textsc{mec} as a function of its volume. In 
the (limited) range of
volumes studied, we find that these lowest-lying excitations are
actually {\it fractal} objects with a dimension $d_f$ close to that of
lattice animals ($d_f=2$); this excitation branch is thus
topologically unrelated to that associated with domain walls.
Furthermore, the exponent $\theta_f$ describing the typical energy of
these fractal \textsc{mec} is measured to be small and negative, 
$\theta_f \simeq - 0.13$.  In contrast, excitations constrained to be compact have
typical energies growing with their size as expected.  Within a
na\"{\i}ve argument, $\theta_f \le 0$ seems incompatible with a
spin-glass ordering at positive temperature.  But such an argument
assumes that excitations of different sizes are
statistically independent, and we find that this
assumption does not hold.

\section{Model and Methods}
We study the $3d$ Edwards-Anderson (EA) model with periodic boundary
conditions.  The Hamiltonian is defined on a cubic lattice of $N=L^3$
spins, \be H =-\sum_{<ij>} J_{ij} S_i S_j \ .
\label{eq_EA}
\ee The spins are Ising, {\it i.e.}, $S_i = \pm 1$, and the
nearest-neighbor interactions $\{J_{ij}\}$ are quenched random
variables distributed according to a Gaussian law with zero mean and
unit variance.

Our measurements are performed on lattices with $N=6^3$ and $N=10^3$
sites; in both cases, we generated $1000$ disorder samples.  For each
disorder sample, we first compute the ground state of the system using
a genetic renormalisation algorithm~\cite{HoudayerMartin99b}.  After,
we choose an arbitrary ``reference'' spin and flip it along with a
cluster containing $v-1$ other spins connected to it.  We then
minimize the energy of this cluster by exchange Monte
Carlo~\cite{HukushimaNemoto95}, but with the constraint that the
reference spin is held flipped and the cluster is always connected and
of size $v$.  Our \textsc{mec} thus differ from the
droplets of Fisher and Huse~\cite{FisherHuse86}.  Indeed, the volume
of a \textsc{mec} is constrained to a fixed value whereas Fisher and Huse
consider the minimum amongst all clusters fitting inside a box of
width $2 \ell$ but not of width $\ell$.  Because of this, the scaling
of energy as a function of characteristic size can very well be
different for these two definitions. However, since all of their
droplets are also \textsc{mec}, droplets so defined actually have
energies below those of our clusters as long as one doesn't
force the droplets to be compact.

To find our \textsc{mec}, we use non-local Kawasaki dynamics
as follows.  First one randomly removes a node of the cluster and then
places it back elsewhere; if the cluster is no longer connected, undo
the change and try again until the modified cluster is connected;
finally, apply the Metropolis condition for accepting or rejecting the
change.  In our exchange Monte Carlo run, we used a total of $35$
temperatures uniformly spaced between $T=0.07$ and $T=2.45$. For each
choice of $v$ ($v_0 = 108, \; v_n=\left\lceil v_0 \times \left({3
\over 4}\right)^n\right\rceil$), we performed runs using different
numbers of Monte Carlo ``sweeps'' where each sweep consists of $v$
accept-reject tests. This allowed us to determine how many sweeps were
necessary to find \textsc{mec} reliably: for the final runs,
we used $10^5$ Monte Carlo sweeps, and then the data at $v\le 33$ is
very reliable.  Unfortunately, at larger values of $v$, the data for
$10^4$ and $10^5$ sometimes disagree so we are not so confident we
have found the optimum there.  Because of this, our analyses are
restricted to $v \le 33$, but for completeness we shall show all of
our data.

\section{Main Results}
Our most striking result is that the average energy $\bar E(v)$ of \textsc{mec}
does not increase with their size $v$ but actually {\it decreases} 
as shown in Fig.~1.
To illustrate the
effects of potential systematic errors, we also show the energy data
from the runs using $10^4$ sweeps only. For the largest $v$, it is
difficult to find the optimum~\footnote{~Note nevertheless that our
average \textsc{mec} energy for $v=108$ is smaller than for $v=3$.}, 
and furthermore the condition that $L$ be much
larger than the cluster's ``extension'' (mean end-to-end distance) is
no longer fulfilled; we thus expect finite size effects to be
significant for $v>33$, and this is corroborated by the measurements
of the cluster's extension and gyration radius.  Because of these two
systematic effects, all of our fits have been performed using the $v
\le 33$ data.

From these mean energies, we extract an estimate of
$\theta_f$. Anticipating that \textsc{mec} have a fractal dimension
$d_f$, we write $\bar E(v) \propto v^{\theta_f/d_f} \propto \ell^{\theta_f}$,
where $\ell$ is the linear dimension of the cluster.  A fit of the
$L=10$ data (with $v\le 33$) to the form $\bar E(v) = A + B/v^{\lambda}$
does not work well and leads to negative values for $A$. On the
contrary, the fit to the form $\bar E(v) = A v^{{\theta_f}/d_f}$ as shown
by a dashed line is very good ($\chi^2\approx 0.7$) and leads to
$\theta_f/d_f = -0.060 \pm 0.006$.

\begin{figure}[htb]
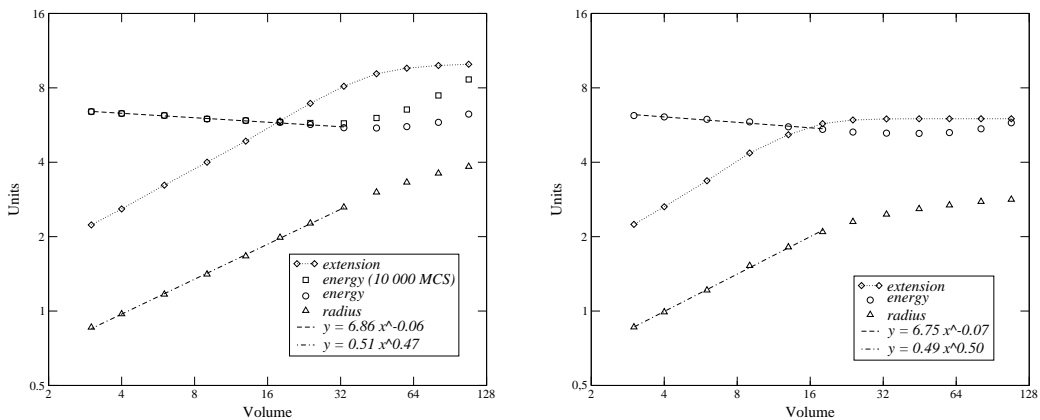

\begin{center}
\includegraphics[width=6.5cm]{fractal10.eps}
\hspace{5mm}
\includegraphics[width=6.5cm]{fractal6.eps}
\caption{Log-log plot of energy, radius of gyration and extension
versus $v$.  Left panel: $N=10^3$ spins; right panel: $N=6^3$ spins.
Also shown is the energy when using $10^4$ rather than $10^5$ Monte
Carlo sweeps. The error bars are smaller than the symbols.}
\end{center}
\label{fig_energies}
\end{figure}

Next, consider the {\it distribution} of excitation energies (see
Fig.~2).
\begin{figure}[htb]
\vspace{7mm}
\begin{center}
\includegraphics[width=7.5cm]{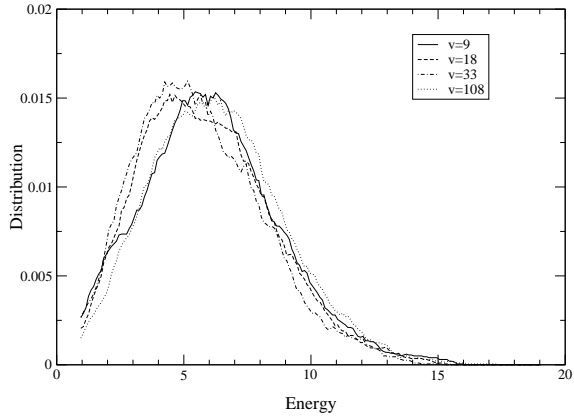}
\caption{Histogram of the excitation energies for the $v=9,18,33,108$
optimal clusters.
For $v=108$, it is interesting to note that the shape of the
distribution hardly changes.  }
\end{center}
\label{fig_histo}
\end{figure}
Surprisingly, the distribution hardly varies at all with $v$. Another
feature is the suppression at energies close to zero~\footnote{~The
hole for small $E$ is observed for all cluster sizes including the
small ones where we are confident that we find the optimum cluster.},
in contrast to what is expected from the droplet model (if we
parametrize these curves by generalized Gamma functions, fits lead to
an exponent close to $2$, {\it i.e.}, $P(E) \simeq E^2 \exp(-\beta
E^2)$).

Before seeing whether these excitations are compact or not, let us now
{\it constrain} them to be so as in the droplet model.  We do this by
forcing the cluster to stay within the cubic box of size $\left\lceil
\gamma v^{1/3} \right\rceil$ with $\gamma=2$, centered around the
reference spin.  Then we find that the energies of these compact
clusters do grow significantly with their volume, in fact as $\bar E(v)
\propto v^{0.19}$.  However the value of this exponent is only indicative
because: (a) the size of the cube takes integer values and therefore
leads to a clear staircase effect; (b) for the relatively small sizes
we considered, the value of the exponent depends on the precise value
of $\gamma$ used. Note that the exponent $0.19$ is larger than the
expected value $\theta/d=0.07$ (with $\theta=0.2$ and $d=3$) 
from the droplet
model. This disagreement may be due to the fact that we work with
fixed volumes; such a constraint leads to 
cluster energies that may grow faster than those of droplets.

Finally, we turn to the geometrical characterization of \textsc{mec}.
Qualitatively, these lowest-energy excitations (in the absence
of any compactness constraint) are quite stringy (see
Fig.~3). 
To quantify this, we have computed the average radius of gyration,
$R_g$, as a function of $v$.  We find that $R_g \propto v^{1/d_f}$ with
$d_f = 2.10 \pm 0.05$ (see Fig.~1). 
This then leads to $\theta_f = -0.13 \pm 0.02$. The same study for our
compact clusters leads to $d_f = 2.7 \pm 0.1$, instead of $3$, but
here again the effects (a) and (b) discussed above lead to important
corrections.

\begin{figure}[htb]
\begin{center}
\includegraphics[width=7cm]{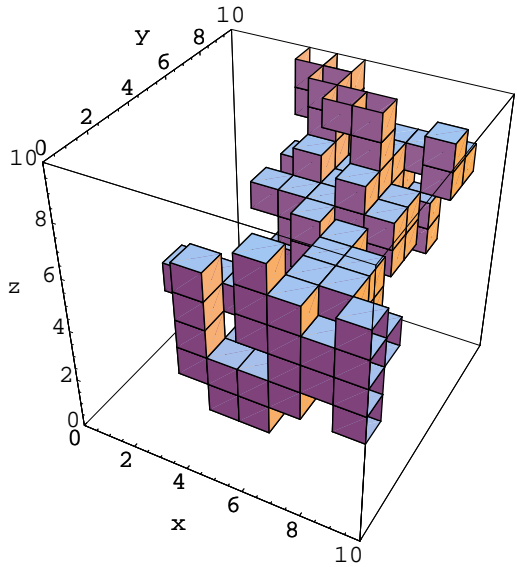}
\caption{A $v=108$ minimal excitation in a $N=10^3$ system
with periodic boundary conditions. }
\end{center}
\label{fig_example}
\end{figure}

By comparing constrained and unconstrained clusters,
we see that compact excitations are {\it not} the ones
of lowest energy.  Actually, a simple entropic argument suggests that
\textsc{mec} may be lattice animals (for which $d_f=2$): the
number of such animals grows exponentially with their volume, 
whereas the number of compact clusters only grows as the exponential of
their surface. Then the non-compact clusters may 
perhaps reach lower energy values simply because
of this greater entropy. To deepen this suggestion, recall
that lattice animals can collapse and become compact when
penalized by a surface energy term~\cite{DH}.  We have added such a 
penalty term to our clusters, allowing us to indeed monitor 
the transition of our {\sc mec} from fractal
to compact.  This and our measurement of $d_f$ suggest that
our \textsc{mec} are in the lattice animal
phase. Further indications come from
geometric quantities such as the
surface to volume ratio or the mass distribution within one
cluster, found to be very similar in both cases (\textsc{mec} and
lattice animals). For example, the ratio 
$\langle r_i^4 \rangle / \langle r_i^2 \rangle^2$, 
where $r_i$ is the distance of site $i$ to
the center of mass of the cluster saturates at a common
value around 1.25 at large $v$.

Our results contradict the main assumptions of the droplet theory: we
find non-compact excitations, decreasing instead of increasing
energies with size, and a hole at zero energy in the probability
distribution of energies.  Such properties are compatible with the
mean field picture, though at this point there have been no studies of
{\it local} excitations in mean field models with finite connectivity.
But one can assert that within such models, all excitations are
non-compact in the sense of having their surface growing as their
volume, and it is plausible that the low-lying
excitations will be the analogs of lattice animals. Note that such a
picture also seems to arise in the limit of both very strong disorder
and high dimensions~\cite{NewmanStein94}.  Finally, these local
excitations may develop smoothly into system-size excitations for
which mean field predicts $\theta=0$.  If this occurs in three
dimensions when the extension of our \textsc{mec} approaches $L$,
then they may turn into the sponge-like system-size excitations found
by Krzakala and Martin~\cite{KrzakalaMartin00a} for which $\theta$ is
also negative or zero.  It would be interesting to develop a new
phenomenology based on fractal droplet excitations and to study its
compatibility with the mean field picture.

\section{Stability of the Spin Glass Phase} 
Very low energy excitations of arbitrarily large size may jeopardize
the stability of the spin glass phase.  In fact, Fisher and
Huse~\cite{FisherHuse86} claim that the stability of the spin-glass
phase requires $\theta>0$.  Their argument assumes that the
\textsc{mec} are compact; when we impose that constraint,
our data are indeed compatible with $\theta >0$, but otherwise we find
$\theta_f \le 0$.  Let us first re-examine their argument in the
context of the droplet model (and thus keeping the
hypothesis of compact droplets). 


They assume that the temperature is very low and that the boundary
conditions force the spins at infinity to take the same values as in
the ground state.  For a given site $i$, consider the order parameter
$m_{i} \equiv\langle S_{i} \rangle$. Its value is 
$m_i =\sigma_i=\pm1$ in the ground state but its magnitude is 
reduced at $T>0$ by the
thermally activated droplets that flip $S_{i}$ with some probability.
Following Fisher and Huse~\cite{FisherHuse86}, let us
neglect the interactions between the droplets; when the temperature
is low, the gas of droplets is expected to be very dilute so the
hypothesis of a non-interacting gas is appropriate.
In this framework, the magnetization at site $i$ 
depends only on the droplets containing that site and is
given by~\cite{FisherHuse86}: $m_{i}(T) = \sigma_i \prod_n \tanh(E_n/2kT)$
where $E_n > 0$ is the excitation energy of droplet $n$ containing
$i$.  Taking $n \to \infty$, $m_{i}$ remains non-zero only if
low-energy droplets are sufficiently rare.  Mathematically, this
reduces to the convergence at large $L$ of the following integral: 
\be
m_{i}(T) \propto \exp \left(-\int_1^L \!  d\Omega_\ell \
\frac{kT}{\Upsilon \ell^\theta} \right) \ ,
\label{qT}
\ee 
where the measure $d\Omega_\ell={d\ell}/{\ell}$ expresses the fact
that the size of droplets must roughly double before the energies
become independent~\footnote{~The formula also assumes that the 
probability distribution of $E/\Upsilon \ell^\theta$ has a non-zero
value at zero argument; the reasoning can be generalized if this
density vanishes, leading to a different dependence on temperature
of $m_i$, but the vanishing or not of $m_i$ in the thermodynamic
limit is unaffected.}.
Now, as long as $\theta >0$, the 
integral over $\ell$ converges, and $m_{i} \ne 0$ in the 
thermodynamic limit~\cite{FisherHuse86}.
Conversely, when $\theta \leq 0$, the probability to excite a droplet
becomes independent of its size $\ell$ for large $\ell$.  One then
finds from Eq.(\ref{qT}) that $m_{i}(T) \propto L^{-\alpha(T)}$, with a
positive exponent $\alpha$ that vanishes for $T \to 0$. Therefore, for
any $T>0$, $m_{i}(T)$ tends to zero for large system sizes.  Our
numerical findings suggest that $\theta_f \le 0$ for {\sc mec}; a
na\"{\i}ve extension of the above argument to {\sc mec} would then suggest
that there is no spin-glass phase, at variance with widely accepted evidence.

However, as we now discuss, two of the assumptions needed to 
obtain (\ref{qT}) are likely not to apply to \textsc{mec}.
First, can \textsc{mec} on different scales be treated as independent?  
One can measure the correlation between 
\textsc{mec} of sizes $v$ and $v'\geq v$
by computing the overlap $q_s(v,v') \equiv \frac{\#(\cal{S} \cap
\cal{S}')}{\#\cal{S}}$, where $\cal{S}$ is the set of surface links
for the excitation of volume $v$, and similarly for $v' > v$. By
definition, one has $q_s(v,v)=1$.  If all the surface links of the
smaller excitation belonged to the surface of the larger one, then one
would also have $q_s(v,v') = 1$.  Our results are plotted in
Fig.~4 
as a function of $v/v'$ for different sizes of
the largest excitation, $v'$.  We find that $q_s(v,v')$ is nearly
independent of $v$ for a given $v'$ and quite large (on the order of
one half) for the sizes investigated. Furthermore, the value of this
``plateau'' only very slowly decreases with $v'$, as a small negative
power or as $1/\log v'$.
\begin{figure}[htb]
\begin{center}
\includegraphics[width=7.5cm]{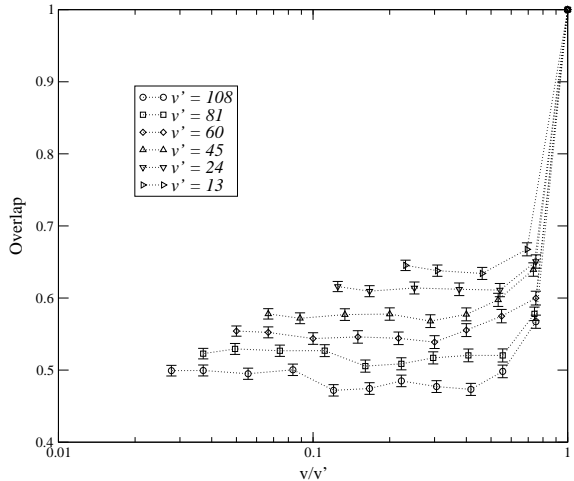}
\caption{Semi-log plot of the normalized intersection $q_s(v,v')$
between surfaces of optimal clusters in the $N=10^3$ system. We show
$q_s$ as a function of $v/v'$ for different largest cluster sizes
$v'$. The ``plateau'' value is seen to decrease slowly with $v'$, as a
small negative power-law or as $1/\log v'$.}
\end{center}
\label{fig_overlap}
\end{figure}
This means that the \textsc{mec} are extremely correlated: an
excitation of size $v$ serves as a good ``backbone'' to construct
larger excitations.  On the other hand, if one had two uncorrelated
fractals, one should observe 
$q_s(v,v') \propto (v')^{1-d/d_f} \propto v'^{-1/2}$, 
{\it i.e.}, a much faster decay than the one seen in
Fig.~4. 
These strong correlations show that our
non-compact \textsc{mec} are not at all independent: the number of
effectively independent clusters is therefore much reduced.  It is not
clear at this stage how the phase space volume $d\Omega_\ell$ has to
be changed to account for these correlations, and whether or not the
resulting integral appearing in (\ref{qT}) would be convergent.  The
slow $1/\log v'$ decay of the correlation function $q_s(v,v')$
suggests that $d\Omega_\ell$ becomes $d \ell/(\ell \log\ell)$, so that
the integral appearing in (\ref{qT}) still diverges, but only as $\log
(\log L)$, leading to $m_i(T) \propto (\log L)^{-\alpha(T)}$ with $\alpha$
vanishing for $T \to 0$. In this case, $m_i(T)$ would therefore be zero
for any $T >0$, but the system size dependence would be so weak for small
temperatures that it would be impossible to disprove this scenario
numerically.

The second hypothesis underlying
(\ref{qT}) is that the \textsc{mec} are dilute enough for the
interaction between two excitations centered around {\it different} sites
can be neglected.  If the integral in (\ref{qT}) diverges for large
$\ell$, then it is no longer self-consistent to neglect 
the interaction between the \textsc{mec}. These interactions 
could then provide a
temperature-dependent cut-off scale when $\theta \le 0$ 
which would make $m_i(T)$ non-zero
for $T > 0$.  This, and the strong correlation between \textsc{mec} of
different sizes, might therefore be enough to reconcile our findings
with the numerical evidence \cite{PalassiniCaracciolo99} of a $T > 0$
spin glass phase for the $3d$ EA model.

\section{Conclusion}
In this paper, we have obtained numerical results on the local
low-lying excitations of the $3d$ Edwards-Anderson model.  We
constructed the optimal connected clusters of flipped spins of a given
size, and studied their energetic and geometrical properties. We find
that these minimum energy clusters (\textsc{mec}) 
have a fractal dimension close to two,
suggesting an analogy with lattice animals.  The energy of these
clusters does not grow with their size (the corresponding energy
exponent is found to be slightly negative).  This is in contrast with
\textsc{mec} constrained to be compact, for which the energy is indeed
found to grow with their size.  We therefore speculate that there
exists a new ``fractal'' excitation branch, that should be included in
an extended droplet theory.  To get some further insights into this
theory, we are now studying these excitations in the presence of a
magnetic field and in four dimensions. Clearly, the static and
dynamical consequences of these objects will have to be worked out.

\section{Acknowledgements}
We thank D. S. Fisher, G. Parisi and J. P. Sethna for many interesting
discussions, D.  A.  Huse for much constructive criticism, and S.
Slijepcevic for participating in an early stage of this work.  J.  L.
acknowledges a fellowship from the MENRT and the CEA for computer time
on the Compaq SC232.

\end{document}